\documentclass{caosp}

\usepackage{graphicx}
\usepackage[utf8]{inputenc}
\usepackage[T1]{fontenc}

\articleNo{0}
\pubyear{0}
\volume{0}
\volnumber{0}
\firstpage{1}
\received{}
\accepted{}

\begin{document}

\title{HT Cas - eclipsing dwarf nova during its superoutburst in 2010 }

\author{
	K.\,B\k{a}kowska \inst{1,} \inst{2}
		\and
		A.\,Olech \inst{1}
		\and
		A.\,Rutkowski \inst{3}
		\and
		R.\,Koff \inst{4}
		\and\\
		E.\,de Miguel \inst{5}
		\and
		M.\,Otulakowska-Hypka \inst{1}
}

\institute{
		Nicolaus Copernicus Astronomical Center, Polish Academy of Sciences\\
		ul. Bartycka 18, 00-716 Warszawa, Poland \\
		\and
		Astronomical Observatory Institute, Faculty of Physics,\\
		A. Mickiewicz University,
		ul. S{\l}oneczna 36, 60-286 Pozna\'{n}, Poland\\
		\and
		Astronomical Observatory, Jagiellonian University, ul. Or{\l}a 171, 30-244 Krak\'{o}w, Poland\\
		\and
		Center for Backyard Astrophysics, Antelope Hills Observatory, 980 Antelope Drive West, 
		Bennett, CO  80102, USA\\
		\and
		Departamento de Fisica Aplicada, Facultad de Ciencias Experimentales, Unversidad de Huelva, 
		21071 Huelva, Spain
}

\date{October 7, 2013}

\maketitle

\begin{abstract}
We present results of a world-wide observing campaign of the eclipsing dwarf nova - HT Cas during its superoutburst in November 2010. Using collected data we were able to conduct analysis of the light curves and we calculated $O-C$ diagrams. 

The CCD photometric observations enabled us to derive the superhump period and with the timings of eclipses the orbital period was calculated. Based on superhump and orbital period estimations the period excess and mass ratio of the system were obtained. 
   
\end{abstract}

\section{Introduction}

\label{intr}

Among close binary stars there are cataclysmic variables a containing white dwarf (the primary) and a main-sequence star (the secondary or the donor). The primary accretes matter from the donor star through the inner Lagrangian point. In non-magnetic systems the material forms an accretion disk around the primary. 

One of the subclasses of cataclysmic variables are dwarf novae and among them there are SU UMa type stars. They can be characterized by short orbital period ($P_{orb} < 2.5$ h) and in their light curves we see two types of outbursts: normal and superoutbursts. Outbursts are about one magnitude fainter and last shorter than superoutbursts. Tooth-shaped oscillations called superhumps manifest their presence during superoutburts (Hellier 2001).

HT Cas was discovered 70 years ago and classified as a variable star (Hoffmeister 1943). Unfortunately, for 35 years this object did not received any attention, until the eclipses of HT Cas were noticed (discovery made by Bond in 1978, private communication with Patterson) and this amazing dwarf nova became a top priority object for observing season organized in 1978.  Patterson called HT Cas "a Rosetta stone among dwarf novae" because of the variety of manifested features presented in light curves (Patterson 1981). Over 30 years since this statement literature concerning HT Cas is still growing, reaching several dozens of publications. 

\section{Observations}

We present observations of superoutburst in HT Cas made during 21 nights. Data were collected between 2nd and 27th of November. During this world-wide campagin five observers were gathering observations in four countries: Poland, Turkey, Spain and USA, and eight telescopes with diameters ranging from 10 to 100 cm were used simultaneously. Moreover, data collected by AAVSO\footnote{American Association of Variable Star Observers, www.aavso.org} organization were used for this analysis.

\begin{figure}
\centerline{\includegraphics[width=5.5cm,clip=]{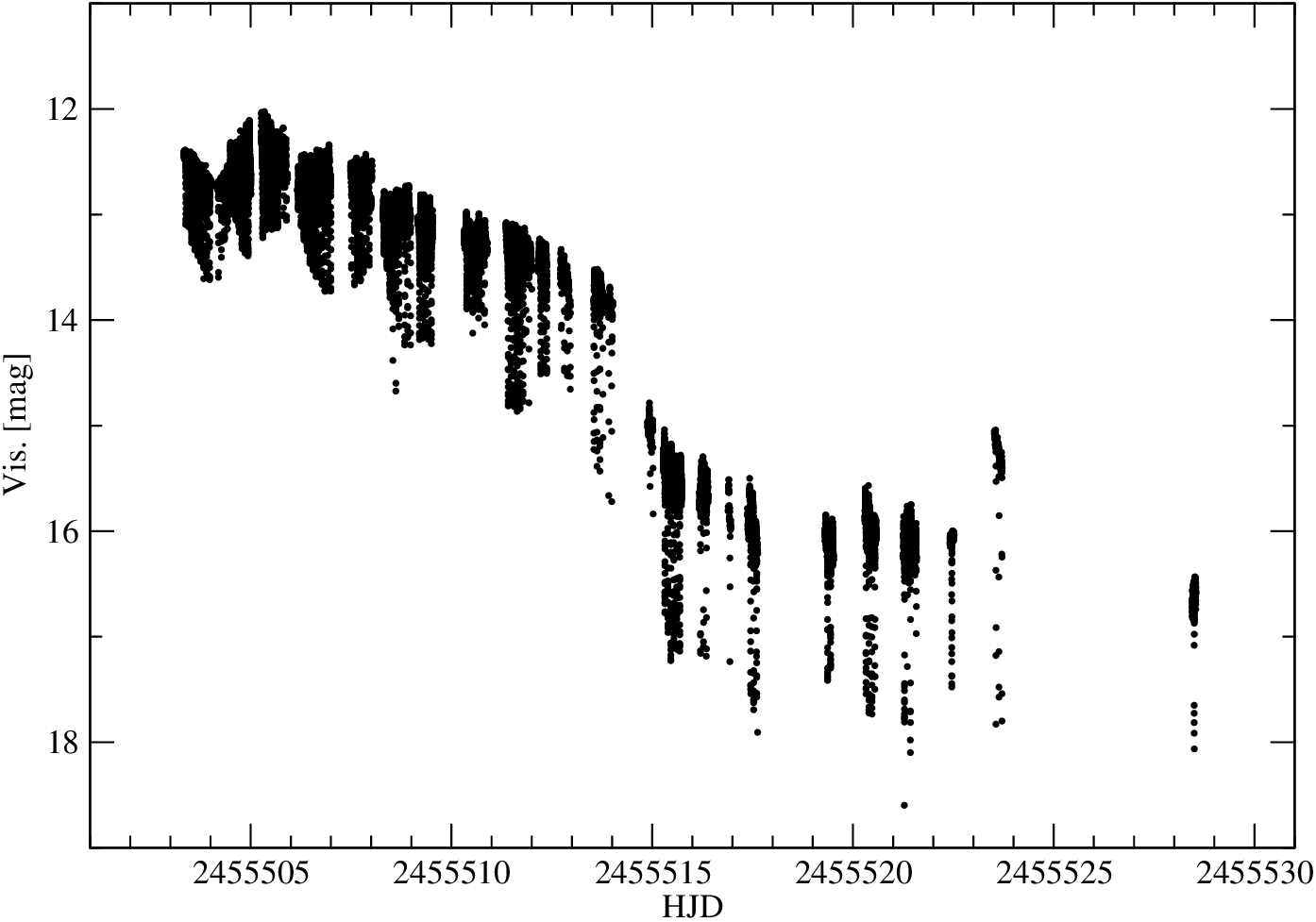}}
\caption{The photometric behavior of HT Cas in November 2010. During that time one can observe an outburst precursor, developed later on into full superoutburst. At the end of November star came back to its  quiescence level. }
\label{f1}
\end{figure}

HT Cas was monitored in clear filter ("white" light). Bias, dark current and flat-field correction was made using IRAF\footnote{IRAF is distributed by
the National Optical Astronomy Observatory, which is operated by the
Association of Universities for Research in Astronomy, Inc., under a
cooperative agreement with the National Science Foundation.} package. Profile photometry was obtained with DAOPHOTII (Stetson 1987). Relative manitudes were transform to the standard Johnson V manitudes using data published by Henden and Honeycutt (1997). In Fig.\,\ref{f1} there are presented resulting light curves from our observing campaign.

The superoutburst lasted 11 nights (3rd-13th Nov) and we gathered data covering almost whole this period except of the last night of the superoutburst. HT Cas reached maximum brightness of $V\approx 12$ mag and the amplitude of this superoutburst was about $A\approx4$ mag.

\section{Results}

The $O-C$ diagram is an excellent tool to check the stability of superhump or orbital period and to determine their values. We analyzed light curves from 10 subsequent nights where superhumps clearly manifested their presents and we identified 69 moments of maxima. The least squares linear fit to the gathered data  gave the following ephemeris for the maxima:
 
\begin{equation}
{\rm HJD_{\rm max}} = 2455504.5132(8) + 0.07608(1) \cdot E
\end{equation} 

\begin{figure}
\centerline{\includegraphics[width=5.5cm,clip=]{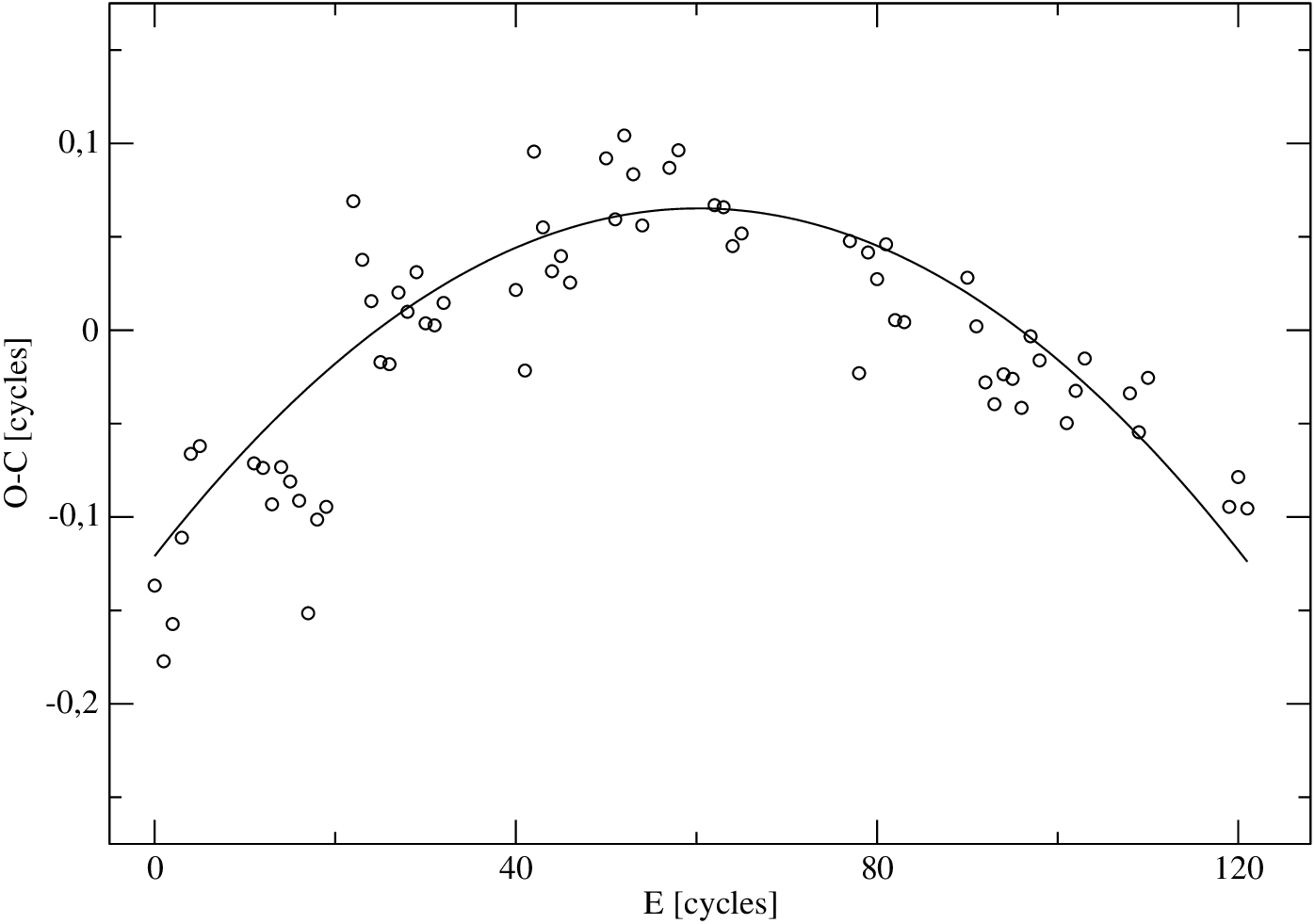}}
\caption{The $O-C$ diagram of the superhumps maxima. The solid line corresponds to the fit given by Eq.\,\ref{eq2}.}
\label{f2}
\end{figure}

One can observe the decreasing trend of superhump period presented in Fig.\,\ref{f2}. Due to this fact the second-order polynomial fit to the moments of maxima was derived and the following ephemeris was obtained:

\begin{equation}
{\rm HJD_{\rm max}} = 2455504.504(1) + 0.07655(6) \cdot E - 3.9(5)\cdot
10^{-6}\cdot E^{2} 
\label{eq2}
\end{equation}

Based on those calculations, we can confirm that the period of superhumps was not stable and can be described by a decreasing trend with a rate of \textit{\.{P}}=$-10.2(1.3) \times 10^{-5}$.

To obtain the value of orbital period we constructed the $O-C$ diagram for the moments of minima. In total the timings of 70 eclipses from November 2010 superoutburst were used to calculate the following ephemeris of the minima:   

\begin{equation}
{\rm HJD_{\rm min}} = 2455504.49185(3) + 0.0736469(5) \times E
\end{equation} 

\begin{figure}
\centerline{\includegraphics[width=5.5cm,clip=]{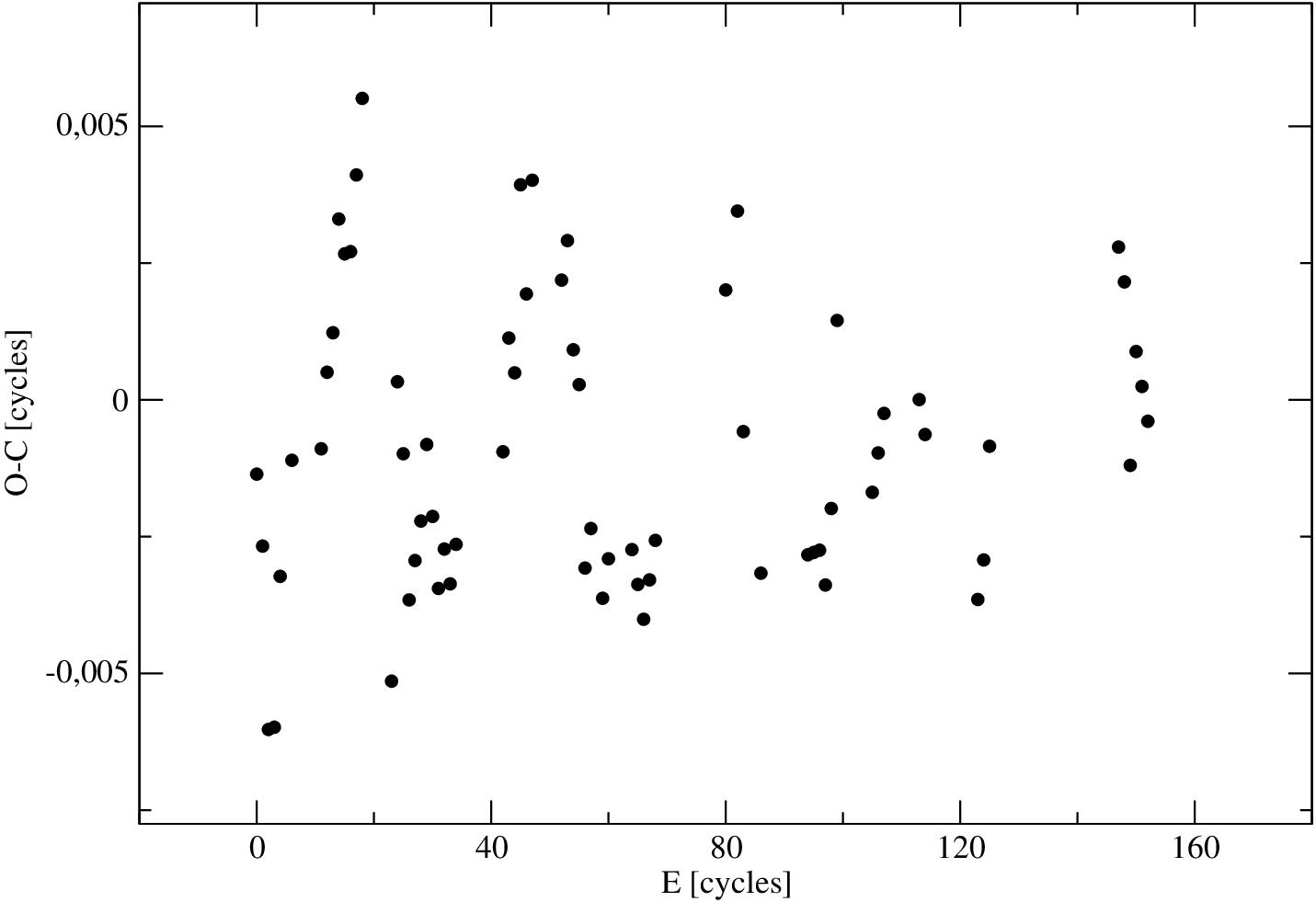}}
\caption{The $O-C$ diagram for  eclipses observed during superoutburst of HT Cas in November 2010. Regular humps are the manifestation of beat period -  composition of the orbital and superhump period.}
\label{f3}
\end{figure}

There is a relation between superhump and orbital period (Osaki 1985):
\begin{equation}
{\frac{1}{P_{sh}} = \frac{1}{P_{orb}} - \frac{1}{P_{beat}}}
\label{eq3}
\end{equation}
and we used Eq.\,\ref{eq3} to calculate the beat period of $P_{beat}=2.30\pm0.01$ days. This value is in full agreement with previous results presented by Zhang et al. (1986) for the superoutburst in HT Cas in 1985. In Fig.\,\ref{f3} one can observe regular humps near cycles: 18, 47 and 80, we checked those features and it is beat period manifestation.

We used values of superhump $P_{sh}$ and orbital $P_{orb}$ periods to calculate the period excess $\varepsilon$ which can be obtained based on formula:

\begin{equation}
\varepsilon = \frac{\Delta P}{P_{orb}} =\frac{P_{sh}-P_{orb}}{P_{orb}}
\label{eq4}
\end{equation}

Based on Eq.\,\ref{eq4} the value of period excess for HT Cas is $\varepsilon=3.30\%\pm0.01\%$ and it is a typical value for SU UMa-type stars. 

There is an empirical formula with the relation between period excess and mass ratio of the binary $q=M_2/M_1$ (Patterson 1998):

\begin{equation}
\varepsilon = \frac{0.23q}{1+0.27q} 
\label{eq5}
\end{equation}

Based on Eq.\,\ref{eq5} the mass ratio of HT Cas with the value of $q=0.149$ was obtained.

\section{Summary}

To summarize the results of the autumn 2010 observations of the HT Cas we can confirm:
\begin{itemize}
\item After 25 years of quiescence or normal outbursts in November 2010 the superoutburst in HT Cas was detected. This rare phenomenon had an amplitude of about $A=4$ mag, lasted 11 nights was triggered by an outburst precursor.   
\item During superoutburst mesmerizing superhumps manifested their presence and based on them the superhump period with a value of $P_{sh}=0.07608(1)$ days and  decreasing with an rate of \textit{\.{P}}$=-10.2(1.3)\times10^{-5}$ was calculated. 
\item Based on the timinings of eclipses observed during superoutburst an orbital period with a value of $P_{orb}=0.0736469(5)$ days was obtained and its value is in full accordance with results presented by other authors from earlier observations (Horne et al. 1991, Feline et al. 1998, Ioannou et al. 1999, Borges et al. 2008,). No anomalies in orbital period were detected as it was mentioned by Ioannou et al. (1999) or Borges et al. (2008). 
\item From the November 2010 superoutburst the period excess with value of $\varepsilon = 3.30\%\pm0.01\%$ was obtained and it has typical value for SU UMa type stars. The same value of period excess was derived from the superoutburst in 1985 (Zhang et al. 1986). 
\item Mass ratio with the value of $q=0.149$ was calculated and it confirms the result obtained by Horne et al. (1991) where a different method and set of observations were used. 

\end{itemize}

\acknowledgements
We acknowledge generous allocation of the Warsaw Observatory and Tzec Maun Foundation for telescope time.  Data from AAVSO observers are also appreciated. The project was supported by Polish National Science Center grants awarded by decisions:   DEC-2012/07/N/ST9/04172 to KB, DEC-2011/03/N/ST9/03289 to MOH, and DEC-2012/04/S/ST9/00021 to AR.

\end{document}